\documentclass[a4paper,11pt]{article}
\usepackage{jheppub} % for details on the use of the package, please
\usepackage[T1]{fontenc} % if needed
\usepackage{graphicx,subfig,epsfig,epsf,color}
\usepackage{amssymb,amsmath}
\usepackage{array}
\usepackage{slashed}
\usepackage{feynmf}
\usepackage{subcaption}
\usepackage{hyperref}
\def\Lda{\Lambda}

\newcommand{\s}{\sigma}

\newcommand{\bea}{\begin{eqnarray}}
\newcommand{\eea}{\end{eqnarray}}
\newcommand{\beas}{\begin{eqnarray*}}
\newcommand{\eeas}{\end{eqnarray*}}
\newcommand{\bcr}{\begin{center}}
\newcommand{\ecr}{\end{center}}
\def\Re{{\cal R \mskip-4mu \lower.1ex \hbox{\it e}\,}}
\def\Im{{\cal I \mskip-5mu \lower.1ex \hbox{\it m}\,}}

\def\etal{{\it et al.}}

\def\tev{\,{\ifmmode\mathrm {TeV}\else TeV\fi}}
\def\gev{\,{\ifmmode\mathrm {GeV}\else GeV\fi}}
\def\mev{\,{\ifmmode\mathrm {MeV}\else MeV\fi}}
\def\to{\rightarrow}

\def\issue(#1,#2,#3){#1 (#3) #2} % PLB format
\def\APP(#1,#2,#3){Acta Phys.\ Polon.\ \issue(#1,#2,#3)}
\def\ARNPS(#1,#2,#3){Ann.\ Rev.\ Nucl.\ Part.\ Sci.\ \issue(#1,#2,#3)}
\def\CPC(#1,#2,#3){comp.\ Phys.\ comm.\ \issue(#1,#2,#3)}
\def\CIP(#1,#2,#3){comput.\ Phys.\ \issue(#1,#2,#3)}
\def\EPJC(#1,#2,#3){Eur.\ Phys.\ J.\ C\ \issue(#1,#2,#3)}
\def\EPJD(#1,#2,#3){Eur.\ Phys.\ J. Direct\ C\ \issue(#1,#2,#3)}
\def\IEEETNS(#1,#2,#3){IEEE Trans.\ Nucl.\ Sci.\ \issue(#1,#2,#3)}
\def\IJMP(#1,#2,#3){Int.\ J.\ Mod.\ Phys. \issue(#1,#2,#3)}
\def\JHEP(#1,#2,#3){J.\ High Energy Physics \issue(#1,#2,#3)}
\def\JPG(#1,#2,#3){J.\ Phys.\ G \issue(#1,#2,#3)}
\def\MPL(#1,#2,#3){Mod.\ Phys.\ Lett.\ \issue(#1,#2,#3)}
\def\NP(#1,#2,#3){Nucl.\ Phys.\ \issue(#1,#2,#3)}
\def\NIM(#1,#2,#3){Nucl.\ Instrum.\ Meth.\ \issue(#1,#2,#3)}
\def\PL(#1,#2,#3){Phys.\ Lett.\ \issue(#1,#2,#3)}
\def\PRD(#1,#2,#3){Phys.\ Rev.\ D \issue(#1,#2,#3)}
\def\PRL(#1,#2,#3){Phys.\ Rev.\ Lett.\ \issue(#1,#2,#3)}
\def\PTP(#1,#2,#3){Progs.\ Theo.\ Phys. \ \issue(#1,#2,#3)}
\def\RMP(#1,#2,#3){Rev.\ Mod.\ Phys.\ \issue(#1,#2,#3)}
\def\SJNP(#1,#2,#3){Sov.\ J. Nucl.\ Phys.\ \issue(#1,#2,#3)}

\title{\boldmath Constraints on light Dark Matter fermions from relic density consideration and Tsallis statistics}

\author[a,1]{Atanu Guha\note{Corresponding author.}}
\author[a]{Prasanta Kumar Das}

\affiliation[a]{Department of Physics, Birla Institute of Technology and Science-Pilani, Goa campus, NH-17B, Zuarinagar, Goa-403726, India }

\emailAdd{p20140401@goa.bits-pilani.ac.in}
\emailAdd{pdas@goa.bits-pilani.ac.in}

\abstract{The cold dark matter fermions with mass MeV scale, pair produced inside the supernova SN1987A core, can freely stream away from the supernovae and hence contributes 
to its energy loss rate. Similar type of DM fermions(having similar kind of coupling to the standard model photon), produced from some other sources earlier, could have contributed to the relic density of the Universe.  Working in a theory with an effective dark 
matter-photon coupling (inversely proportional to the scale $\Lambda$) in the formalism 
of Tsallis statistics, we find the dark matter contribution to the relic 
density and obtain a upper bound on $\Lda$ using the experimental bound 
on the relic density for cold non-baryonic matter i.e. $\Omega h^2 = 0.1186 \pm 0.0020 $. 
The upper bound obtained from the relic density is shown with the lower bound obtained 
from the Raffelt's criterion on the emissibity rate of the supernovae SN1987A energy loss
$\dot{\varepsilon}(e^+ e^- \to \chi \overline{\chi}) \le 10^{19}~\rm{erg~g^{-1}s^{-1}}$  
and the optical depth criteria on the free streaming of the dark matter 
fermion (produced inside the supernovae core). As the deformation parameter $q$ changes 
from $1.0$ (undeformed scenario) to $1.1$(deformed scenario), the relic density 
bound on $\Lda$ is found to vary from $ \sim 4.9 \times 10^7 $ TeV to $1.6 \times 10^8$ TeV 
for a fermion dark matter($\chi$) of mass $m_\chi = 30~\rm{MeV}$, which is almost $10$ times more than the lower bound obtained from 
the SN1987A energy loss rate and the optical depth criteria.
 }

\begin{document} 
\maketitle
\flushbottom

\section{Introduction}
\noindent Experimental evidences from DAMA-LIBRA, CRESST, SuperCDMS, LUX, PICO are 
strengthening the concept of dark matter(DM) day by day. It is now well established 
that dark matter is essential to build the large-scale structure of our Uniniverse. At galactic and 
sub-galactic scales, there are evidences for the structure formation. These includes 
galactic rotation curves, the weak gravitational lensing of distant galaxies by 
foreground structures. In 1932 the Dutch astronomer Jan Hendrik Oort analyzed the 
acceleration of matter by studying the vertical motions of all 
known stars near the Galactic plane, which can be thought as the first indication 
of some unseen mass. After estimating the gravitational potential for the luminous matter Oort 
surprisingly found out that the potential necessary to keep the known stars bound to the 
Galactic disk is simply not sufficient\cite{MRoos:2010wb, MRoos:2012cc}. 
The second possible indication (historically) for the possible presence of dark matter at a 
cosmological  distance scale in our galaxy, was found in 1933 by Fritz Zwicky \cite{Zwicky}. 
 Zwicky measured the radial velocities of member galaxies in the Coma cluster and the cluster radius. 
Using the virial theorem Zwicky made an estimate of the average mass 
of the galaxies within the cluster and found that it is 160 times larger than expected 
from their luminosity and from this he proposed that the missing matter was dark.  He 
found that the orbital velocities of member galaxies in the Coma cluster were almost a 
factor of ten larger than expected from the summed mass of all galaxies belonging to the 
Coma cluster. A large amount of non-luminous matter, dubbed as dark matter, is required 
in order to hold galaxies together the cluster. Current data which constrain the energy densities of the
Universe in normal matter (baryons), dark matter and dark energy $\Lambda$ to
be $\sim 5\%$, $\sim 27\%$ and $\sim 68\%$, respectively. This means that the normal 
matter  we know and that makes up all stars and galaxies only accounts for $5\%$ of 
the content of the universe! Dark matter, five times more than the normal luminous
matter, accounts for a quarter of the Universe \cite{Feng}. 

\noindent Since the DM  has no electric or magnetic charge, it does not interact 
electromagnetically with the normal luminous matter; even if it does so, the 
interaction is very weak. It does not absorb, reflect or emit light, making it 
 extremely hard to detect.  So far, researchers have been able to infer the 
 existence of dark matter only from the gravitational effect it seems to have on 
 visible matter. \\
 But the nature of dark matter remains a mystery for a long time. 
 A wide range of collider and astrophysical study suggests that it is a Weakly 
 Interacting Massive Particle(WIMP) of mass ranging from a few MeV to few tens of 
 GeV. Theories suggest that DM candidates are most likely to be found in the 
 beyond the Standard Model(SM) physics e.g. in models with supersymmetry or extra dimension(s) etc. 
 Direct detection of DM includes its interaction with nucleons 
in underground detectors, whereas indirect detection through DM annihilation to SM states
(i.e. neutrinos) inside the Sun  has been done.
%, in the Galactic center (to photons), and in the Galactic neighborhood (to charged particles).
Experiments at the Large Hadron Collider(LHC) and the 
upcoming electron-positron linear collider(LC) will give more information about the dark matter as the  
missing energy signature. See \cite{JKG,BHS,Feng} for a review on dark matter searches.
\noindent Here we are to investigate the light dark matter fermions contribution to relic density. The concerned
dark matter fermions may be pair produced in the crust of the supernovae core, which afterwords can 
freely stream away while taking away the energy released in supernovae explosion and also from some other sources like bullet cluster etc. In a work
Guha \etal \cite{Atanu} investigated the role of fermion dark matter in the supernovae 
SN1987A cooling: they worked in an effective DM model where the SM photon couples with 
the dark matter fermion through a magnetic/electric dipole moment operator. Working with the formalism 
of Tsallis statistics and applying the Raffelt's criteria \cite{Raffelt} on the supernovae energy loss rate and free streaming 
criteria, they found a lower bound on $\Lda \sim 10^6 - 10^7$ TeV. The DM produced inside the supernovae crust may contribute to the relic 
density after they freely stream away from the supernovae. Here we are to investigate the DM 
relic density upper bound on $\Lda$ and show those along with the lower bounds obtained from the SN1987A 
energy loss rate and free streaming criteria on DM fermions. \\
~~The outline is as follows. In Sec. II, we give a brief 
description of the relic density calculation and introduce the Tsallis statistics (characterized 
by the deformation parameter $q$). In Sec. III, we discuss the contribution of similar type of DM fermions to the relic density, which is pair produced in electron-positron annihilation inside the SN1987A 
core. But they could have been produced from some other sources as well, only thing is that, they are similar in nature and they couple to the standard model photons in a similar way. We obtained their contribution to the relic density using the lower bound on $\Lda$ obtained previously from SN cooling and free streaming \cite{Atanu}. In that way this contribution signifies the minimum contribution of the concerned DM fermions. Because relic density contribution is directly proportional to the scale $\Lda$. Also we can see the contribution is very less(almost $0.1\%-1\%$ of the total non-baryonic density, $0.1186$). This is because they are very light(low mass contribution) and if we consider the supernovae explosions as one of the major production process of those DM fermions, the explosion energy is not enough to produce them in significant amount. Consequently they can not significantly contribute to the dark matter relic density. But we can obtain a lower and upper bound on the effective  scale $\Lda$ from these two consideration(SN cooling and relic density).
The numerical analysis part is presented in Sec. IV. Using the experimental 
value of the non-baryonic relic density i.e. $0.1186 \pm 0.0020$ (obtained from 
the measurement of CMB(cosmic microwave background) anisotropy and the spatial 
distribution of galaxies), we obtain a upper bound on the scale $\Lda$ of the dark 
matter effective theory in the deformed ($q > 1$) and undeformed ($q=1$) 
scenarios, respectively. Using the Raffelt's criteria and the optical depth criteria(based on free streaming 
of dark matter fermions), we obtained the lower bound on $\Lda$ \cite{Atanu} and in the present work we show them with the relic density bound obtained in Sec. IV. Finally, we summarize our results 
and conclude in Sec. V.

\section{Boltzmann equation, Relic density calculation and $q$-deformed statistics}
\subsection{Relic density contribution and experimental estimation}

 The time evolution of the phase space distribution in plasma cosmology is described by Liouville equation, where we consider the dynamics of plasma(ionized gas) as a key element in describing the physics of the large-scale structure formation of the Universe \cite{kolb}.
 \begin{flushleft}
{\bf Liouville equation} 
\end{flushleft}
\bea
\frac{d \rho}{dt}= \frac{\partial \rho}{\partial t} + \sum_{i=1}^{n} \left( \frac{\partial \rho}{\partial q_i} \dot{q_i} + \frac{\partial \rho}{\partial p_i} \dot{p_i} \right)
\eea
\noindent An era of the early stages of the Universe, when various particle candidates fall out of thermal equilibrium with each other, is popularly known as freeze-out. Rapid expansion of the Universe at that time is mainly responsible for this which causes the interaction rate of those particle to decrease. As a result they don't interact to each other further and tend to contribute to the cosmic abundances as a form of mass of radiation. After freeze-out(decoupling) the microscopic evolution of the phase space distribution of a particle species is described by Boltzmann equation \cite{kolb, gondolo}
  \bea
L[f]=C[f] 
 \eea
where, $L$ is the Liouville operator and $C$ stands for the collision operator which represents the number of concerned particles per unit phase-space volume those are lost or gained per unit time under collision with other particles.

\noindent Relativistic generalized form of the Liouville operator becomes
\bea
\hat{L}= p^{\alpha} \frac{\partial}{\partial x^{\alpha}} - \Gamma^{\alpha}_{\beta \gamma} p^{\beta} p^{\gamma} \frac{\partial}{\partial p^{\alpha}}
\eea 

\noindent For spatially homogeneous and isotropic phase-space density the Liouville operator in the Friedmann-Robertson-Walker cosmological model is given by
\bea
L[f]=\frac{\partial f}{\partial t}- H \frac{\mid p \mid^2}{E} \frac{\partial f}{\partial E}
\eea 

\noindent Therefore the Boltzmann equation in FRW cosmology becomes
\bea
\frac{dn}{dt}+3Hn=\frac{g}{(2 \pi)^3} \int \hat{C}[f] \frac{d^3 p}{E} 
\eea 
where, the number density of the particle in terms of phase space density is given by
\bea
n(t)=\frac{g}{(2 \pi)^3} \int d^3 p f(E,t)
\eea
and $g$ stands for the degree of freedom.

\noindent Defining two new variables $Y=\frac{n}{s},~x=\frac{m}{T}$($s, m, T$ denotes the entropy density, mass of the particle species and temperature respectively) and writing down the collision term for a particular process we get the Boltzmann equation as follows \cite{kolb, gondolo, majumdar}
\bea
\frac{dY}{dx}=\frac{1}{3H} \frac{ds}{dx} <\sigma v_{rel}> \left(Y^2 - Y^2_{eq} \right)
\eea
with the thermal averaged crosssection times relative velocity
\bea
<\sigma v_{rel}>=\frac{\int \sigma v_{rel} e^{-E_1/T} e^{-E_2/T} d^3 p_1 d^3 p_2}{\int e^{-E_1/T} e^{-E_2/T} d^3 p_1 d^3 p_2}
\label{eq:sigmav}
\eea

\noindent Working in the standard FRW cosmology 
\bea
H= \left(\frac{8}{3} \pi G \rho \right)^{\frac{1}{2}}
\eea 
we can substitute the following expressions for the density $\rho$ and entropy density $s$ in Eq.(\ref{eq:sigmav})
\bea
\rho= g_{eff}(T) \frac{\pi^2}{30} T^4 \nonumber \\
s=h_{eff}(T) \frac{2 \pi^2}{45} T^3
\eea
and subsequently we get the following form of the Boltzmann equation \cite{gondolo, majumdar}
\bea
\frac{dY}{dx}=-\left(\frac{45}{\pi} G \right)^{-\frac{1}{2}} \frac{m}{x^2} g_{*}^{\frac{1}{2}} <\sigma v> \left(Y^2- Y^2_{eq} \right)
\eea
where,
\bea
g_{*}^{\frac{1}{2}}= \frac{h_{eff}}{g_{eff}^{\frac{1}{2}}} \left( 1+ \frac{1}{3} \frac{T}{h_{eff}} \frac{d h_{eff}}{dT}\right)
\eea
 with the total effective degree of freedom for all final species $g_{eff}(T)=\sum_i g_i(T)$ and $h_{eff}(T)=\sum_i h_i(T)$. The effective degrees of freedom for each species
are given by \cite{gondolo}
\bea
g_i(T)= \frac{15 g_i}{\pi^4} x_i^4 \int_{1}^{\infty} \frac{z \sqrt{z^2-1}}{\exp(x_i z)+\eta_i} z dz \nonumber \\
h_i(T)=\frac{45 g_i}{4 \pi^4} x_i^4 \int_{1}^{\infty} \frac{z \sqrt{z^2-1}}{\exp(x_i z)+\eta_i} \frac{4z^2-1}{3z} dz
\eea
with $x_i=m_i/T$, where, $m_i$ stands for the mass of that particular species and $\eta_i=1$ for Fermi-Dirac statistics and $\eta_i=-1$ for Bose-Einstein statistics.

\noindent After decoupling(freezing-out) we can neglect the term $Y_{eq}$ \cite{majumdar} and integrating from the freeze-out period to the present epoch we get
\bea
\frac{1}{Y_0}=\left(\frac{45}{\pi} G \right)^{-\frac{1}{2}} \int_{T_0}^{T_f} g_{*}^{\frac{1}{2}} <\sigma v> dT
\label{eqn:Y0}
\eea
Again the value of $\frac{1}{Y}$ at $T=T_f$ has been neglected as at freeze-out the number density of the concerned particle is considered to be very high which results the term $\frac{1}{Y_f}=\frac{s_f}{n_f}$ to be very small compared to the other terms. 

\noindent Relic density of the concerned particle at present  can easily be obtained after evaluating $Y_0$ in the units of the critical density as follows \cite{edsjo, gelmini}
\bea
\Omega_{\chi}=\frac{\rho^0_{\chi}}{\rho_{crit}}=\frac{m_{\chi} s_0 Y_0}{\rho_{crit}}
\eea 
with the critical density $\rho_{crit}=\frac{3H^2}{8 \pi G}$, $s_0$ in the entropy density today. With the knowledge of the present day background radiation temperature $T_0=2.726~\rm{K}=2.35 \times 10^{-16}~\rm{TeV}$ we obtain
\bea
\Omega_{\chi} h^2=2.755 \times 10^{11}~m_{\chi}~Y_0
\label{eqn:relic_density}
\eea

Where the value of $m_{\chi}$ is in TeV.
The experimental value of the non-baryonic relic density i.e., $0.1186 \pm 0.0020$, has been obtained from the measurement of CMB(cosmic microwave background) anisotropy and the spatial distribution of galaxies (PDG 2017).

\subsection{Fluctuating temperature and Tsallis statistics}
  $\chi^2$ distribution takes the following form  in the $q$-deformed statistics \cite{Beck_cohen} to account for the temperature ($T$) fluctuations 
\cite{Kaniadakis} 
\bea
f(\beta) = \frac{1}{\Gamma\left(\frac{n}{2}\right)} \left(\frac{n}{2 \beta_0} \right)^{n/2} \beta^{\frac{n}{2} - 1} ~\exp\left(- \frac{n \beta}{2 \beta_0}\right)
\label{fbeta}
\eea 
where $n$ is the degree of the distribution and $\beta = {1 \over {k T}}$. 
%For $n$ independent Gaussian random variables $X_i,~i = 1,....,n$, one can define 
The average of the fluctuating inverse temperature $\beta$ can be estimated as 
\bea
\langle \beta \rangle = n \langle X_i^2 \rangle = \int_{0}^{\infty} \beta f(\beta) d\beta = \beta_0
\eea 
Taking into account the local temperature fluctuation, integrating over all $\beta$, we find
the $q$-generalized relativistic( with particle energy $E = \sqrt{{\bf p}^2 c^2 + m^2 c^4}$) 
Maxwell-Boltzmann distribution 
\bea
{\mathcal{P}}(E) \sim  \frac{E^2}{\left(1 + b(q-1)E\right)^{\frac{1}{q-1}}}
\eea
where  $q = 1 + \frac{2}{n + 6}$ and $b = \frac{\beta_0}{4 - 3 q}$. Its generalization to Fermi-Dirac 
and Bose-Einstein distribution is worked out in \cite{Beck_super}.  The average occupation number of 
any particle within this $q$-deformed statistics ( Tsallis statistics \cite{Tsallis}) formalism is given by  $f_i(\beta,E_i)$ ($i = 1,2$ corresponds to particles) where
\bea
f_i(\beta,E_i) = \frac{1}{\left(1 + (q-1)b E_i \right)^{\frac{1}{q-1}} \pm 1}
\eea
where the $-$ sign is for bosons and the $+$ sign is for fermions. 
Note that the effective Boltzmann factor 
$x_i = \left(1 + (q-1)b E_i \right)^{-\frac{1}{q-1}}$ approaches to the ordinary Boltzmann factor 
$e^{- b E_i} (= e^{- \beta_0 E_i})$ as $q \to 1$.   
 For more discussion related to this and to find its various applications please refer to \cite{Atanu, Atanu3, Atanu4, Tsallis_book, Sumiyoshi_book, Beck_cosmic, Bediaga-ep, Beck-ep, Beck_deform, Beck_dark}.
 
\subsection{Brief discussion on SN1987A cooling and Free streaming of produced DM particles} 

\subsubsection{SN1987A cooling}
The supernova SN1987A was the most evident example of a core-collapse type II supernova explosion till date, which was even visible to the naked eye. After four days of the SN1987A (or AAVSO 0534-69) explosion in the Large Magellanic Cloud (a dwarf galaxy satellite of the Milky Way), a blue supergiant massive star was disappeared. Thus the progenitor of SN1987A was identified as Sanduleak ($M \sim 20\,M_\odot$), a B3 supergiant in the constellation Dorado at a distance approximately $51.4$ kiloparsecs ($1.68\times 10^5$ light-years) from Earth. An enormous amount of energy was released in the SN1987A explosion which equals to the gravitational binding energy $E_g$ of the proto-neutron star (of mass $M_{PNS}$) which is given by
\bea
E_g = \frac{3 G_N M_{PNS}^2}{5 R_{NS}} \sim 3.0 \times 10^{53} \;{\rm erg.}
\eea
Here $M_{PNS} = 1.5 M_\odot$, $R_{NS} = 10~{\rm Km}$ and $G_N$ is the Newton's gravitational constant. 
As per present understanding, neutrinos carry away $99\%$ of the huge amount of released energy and the remaining $1\%$ contributes to the kinetic energy of the explosion. For the earth based detectors the primary astrophysical interest was to detect this neutrino burst. This neutrino flux was first detected by the two collaborations Kamiokande \cite{Kamioka} and IMB\cite{IMB} using their earth based detectors.
The data obtained by them suggest that, in a couple of seconds about 
$10^{53}~\rm{ergs}$ energy was released in the SN1987A explosion.  
The observed neutrino luminosity in the detector(IMB or Kamiokande) 
is $L_\nu \sim 3 \times 10^{53}\;{\rm erg ~s^{-1}}$ (including $3$ generations of neutrinos and anti-neutrinos 
i.e. $\nu_e,~\nu_\mu,~\nu_\tau$ and ${\overline{\nu_e}},~{\overline{\nu_\mu}},~{\overline{\nu_\tau}}$). 
So $\tilde{L}_\nu = \frac{L_\nu}{6} \sim 3 \times 10^{52}~{\rm erg~ s^{-1}}$. The mass of a typical 
proto-neutron star $M_{PNS} = 1.5 M_\odot = 3 \times 10^{33}~{\rm g}$. So, the average energy loss per unit mass is $\frac{\tilde{L}_\nu}{M_{PNS}} \simeq 1 \times 10^{19}~{\rm erg ~g^{-1}s^{-1} }$. 
Note that this is the energy carried away by each of the above 6 (anti)-neutrino species. 
 
For any new physics channel, Raffelt's criteria states that, besides neutrino, if Kaluza-Klein graviton, Kaluza-Klein radion, axion also take away energy, 
the energy-loss rate due to these new channels $\epsilon_{new}$ should be less than the above average energy 
loss rate \cite{Raffelt}, i.e. 
\bea
\epsilon_{new} \le 10^{19}~\rm{erg~g^{-1}~s^{-1}}
\eea
and this follows from the observed neutrino luminosity per species (total six neutrinos and anti-neutrinos, three types each). 
If any energy-loss mechanism has an emissivity greater than $10^{19}~\rm{erg~g^{-1}~s^{-1}}$, then it will remove 
sufficient energy from the explosion to invalidate the current understanding of core-collapse supernova.

Using the Raffelt's criteria of  the supernova energy loss rate for any new physics channel, we constrained the scale $\Lda$ of the dark matter effective theory \cite{Atanu}. Now in a realistic scenario, since the core temperature of the supernova is fluctuating,  we worked within the formalism of Tsallis statistics\cite{Tsallis, Atanu, Atanu3, Atanu4} where this temperature fluctuation is taken into 
account. 

\subsubsection{Free streaming of produced dark matter fermions from SN1987A}

Now the constraint on the effective scale $\Lambda$ of the dark matter effective theory obtained using the Raffelt's criteria 
holds to be truly sensible if the produced dark matter fermion free streams out of the supernova without getting trapped. In order to find the free streaming/trapping their mean free path is to be evaluated \cite{Dreiner}, which is given by
\bea
\lambda_{\chi}=\frac{1}{n_e \cdot \sigma_{e \chi \rightarrow e \chi}}
\protect\label{mean_free_path}
\eea
where $n_e(= 8.7 \times 10^{43}~\rm{m^{-3}})$ is the number density of the colliding electrons in the supernova 
and $ \sigma_{e \chi \rightarrow e \chi}$ is the cross section for the scattering of 
the dark matter fermion on the electron which is related via the crossing symmetry to the annihilation 
cross section $\sigma_{e e  \rightarrow \chi  \chi}$.  
Now, most of the dark matter particles produced in the outermost $ 10\% $ of the star ($ 0.9 R_{c} < r < R_{c} $) from 
electron-positron annihilation  \cite{DHLP, Atanu}. Then any of the dark matter 
particles produced in electron-positron annihilation while propagating through the proto-neutron star, 
can undergo scattering due to the presence of neutrons and electrons inside the star. 
In the case of supernova cooling, neutron-dark matter particle scattering will be negligible 
for free streaming due to neutron mass \cite{Dreiner, Atanu}. We use the optical depth criteria \cite{Dreiner} 
%[arXiv:hep-ph/0304289]
\bea
\int_{r_0}^{R_c}\frac{dr}{\lambda_{\chi}} \leq \frac{2}{3}
\eea
to investigate whether the dark matter fermion produced at a depth $r_0$ free streams out of the supernova 
and takes away the released energy or gets trapped inside the supernova. Here we set 
$r_0=0.9 R_c$ in our analysis, where $R_c $ ($ \simeq 10 $ km) is the radius of the supernova core 
(proto-neutron star) \cite{Dreiner, Atanu}. From the optical depth criteria, we find that 
minimum length of the mean free path for free streaming $\lambda_{fs}$ is $\lambda^{\rm{min}}_{fs} = 1.5~\rm{km}$ 
and it increases with $\Lambda$. On the other hand we can get a lower bound on the effective scale $\Lambda$ using the optical depth criterion which has been done in the numerical section and also in \cite{Atanu}.

%%%%%%%%%%%%%%%%%%%%%%%%%%%%%%%%%%%%%%%%%%%%%%%%%%%%%%%%%%%%%%%%%%%%%%%%%%%%%%%%%%%%
\section{Light fermionic Dark matter contribution to the relic density}
\begin{figure}[!ht]
  \centering
  \subfloat[$ e^-e^+ \stackrel{\gamma}{\longrightarrow} \chi \bar{\chi}$]{\includegraphics[width=0.49\textwidth]{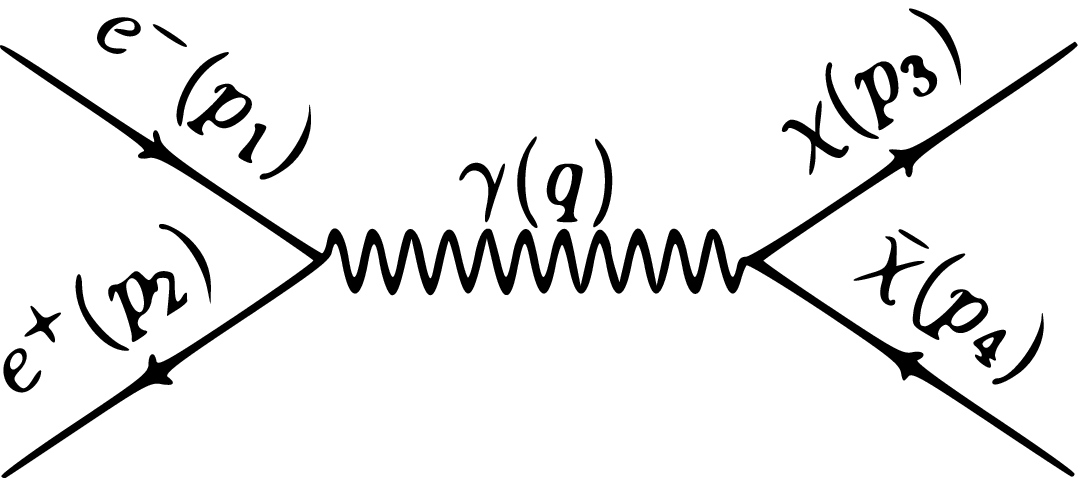}\label{feynman1}}
  \hfill
  \subfloat[$ \chi \bar{\chi} \stackrel{\gamma}{\longrightarrow} e^-e^+$]{\includegraphics[width=0.49\textwidth]{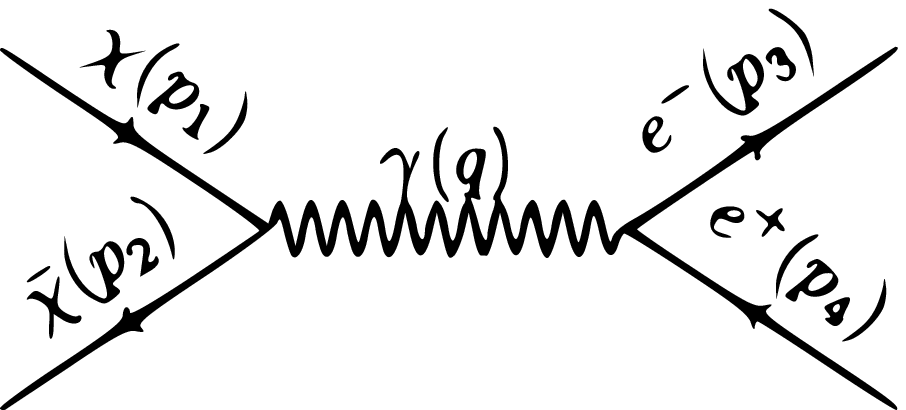}\label{feynman2}}
   \caption{\it Feynman diagram for the processes $ e^-e^+ \stackrel{\gamma}{\longrightarrow} \chi \bar{\chi}$ and $ \chi \bar{\chi} \stackrel{\gamma}{\longrightarrow} e^-e^+$.}
\label{feynman}
\end{figure}

Electrons are abundant in supernovae. The dark matter fermions may be pair produced in the $s$-channel 
annihilation of electron and positron [Fig.~\ref{feynman1}]: $ e^-(p_1)e^+(p_2) \stackrel{\gamma}{\longrightarrow} \chi(p_3)
\bar{\chi}(p_4) $.
%%%%%%%%%%%%%%%%%%%
%%%%%%%%%%%%%%%%%%%%%%%%%%%%%%%%%%%%%%%
%%%%%%%%%%%%%%%%%%%%%%%%%%%%%%%%%%%%%%%%
 To estimate the contribution to the relic density of the produced dark matter fermion, we have to look for the reverse process [Fig.~\ref{feynman2}] $ \chi \bar{\chi} \stackrel{\gamma}{\longrightarrow} e^-e^+ $.
%%%%%%%%%%%%%%%%%%%
%\tikzset{
%photon/.style={decorate, decoration={snake,amplitude=4pt, segment length=5pt}, draw=black},
%particle/.style={draw=black, postaction={decorate}, decoration={markings,mark=at position .5 with {\arrow[draw=black]{>}}}},
%antiparticle/.style={draw=black, postaction={decorate}, decoration={markings,mark=at position .5 with {\arrow[draw=black]{<}}}}
%,gluon/.style={decorate, draw=black, decoration={coil,amplitude=4pt, segment length=5pt}}
%}
%\begin{center}
% \begin{tikzpicture}[thick,scale=1.0]
%\draw[particle] (-1.5,1) -- node[black,above,sloped] {$ \chi(p_1) $} (0,0);
%\draw[antiparticle] (-1.5,-1) -- node[black,above,sloped] {$ \bar{\chi} (p_2) $} (0,0);
%\draw[photon] (0,0) -- node[black,above,sloped] {$ \gamma (q) $} (2,0);
%\draw[particle] (2,0) -- node[black,above,sloped] {$ e^- (p_3) $} (3.5,1);
%\draw[antiparticle] (2,0) -- node[black,above,sloped] {$ e^+ (p_4) $} (3.5,-1);
%\end{tikzpicture}
%\caption{{FIG.0: \it Feynman diagram for the process $ \chi
%\bar{\chi} \stackrel{\gamma}{\longrightarrow} e^-e^+$}}
%\end{center}
%%%%%%%%%%%%%%%%%
%\vspace*{-0.15in}

 The effective Lagrangian of describing photon($\gamma$) and dark matter fermion ($\chi $) interaction is given 
by
\bea 
\mathcal{L}=-\frac{i}{2}\bar{\chi}\sigma_{\mu \nu}(\mu_{\chi}+\gamma_5 d_{\chi})\chi F^{\mu \nu}
\eea
\noindent where $F^{\mu\nu} = \partial^\mu A^\nu - \partial^\nu A^\mu$, the electromagnetic field strength tensor. 
Here $ \mu_{\chi} $ and $ d_{\chi} $ correspond to the magnetic dipole moment and the electric dipole 
moment  of the dark matter fermion $\chi$. $\sigma^{\mu \nu}=\frac{i}{2}[\gamma^\mu,\gamma^\nu] $ is 
the spin tensor. The $4$-momentum vectors of the initial and final state particles (Fig. \ref{feynman2}) in the center-of-mass
frame are given by
\begin{eqnarray}
p_{1} &=&  \left(E, 0, 0, p_z \right); ~p_{2} = \left(E, 0, 0,-p_z \right); \nonumber \\  
%\end{equation}
%\begin{equation}
p_{3} &=& \left(E^\prime, p \sin \theta \cos \phi, p \sin \theta \sin \phi, p \cos \theta \right);\nonumber \\
%\end{equation}
%\begin{equation}
p_{4} &=& \left(E^\prime,-p \sin \theta \cos \phi,- p \sin \theta \sin \phi,-p \cos \theta \right). \nonumber 
\end{eqnarray}  
%Feynman Rules:Vertex factor for $ e^-e^+\rightarrow \gamma $ vertex:$ ie\gamma^{\mu} $
%Vertex factor for $ \gamma\rightarrow \chi \bar{\chi} $ vertex:
%$ i(\mu_{\chi}\sigma^{\mu \nu} q_\nu + d_{\chi}\sigma^{\mu\nu}q_\nu\gamma^5) $ 
%\bea \label{eqn:Spin averaged Amplitude Square_Trace form}
%\overline{|\mathcal{M}|^2}=\frac{e^2}{4q^4}Tr[\slashed{p_1}\gamma^\alpha\slashed{p_2}\gamma^\mu] \cdot Tr2 
%\eea
%\noindent
%where, Tr2=$Tr[(\slashed{p_3}+m_\chi)(\mu_{\chi}\sigma_{\mu \nu} q^\nu + d_{\chi}\sigma_{\mu \nu}q^\nu\gamma^5)(\slashed{p_4}-m_\chi)(\mu_{\chi}\sigma_{\alpha \beta} q^\beta - d_{\chi}\sigma_{\alpha \beta}q^\beta\gamma^5)]$ 
The spin-averaged amplitude square for the process $ \chi(p_1)
\bar{\chi}(p_2)  \stackrel{\gamma}{\longrightarrow} e^-(p_3)e^+(p_4) $ is given by 
\bea \label{eqn:Spin averaged Amplitude Square}
\overline{|\mathcal{M}|^2}=4 \pi \alpha[\mu_\chi^2\left\lbrace s (1-\cos^2 \theta)+4m_\chi^2(1+\cos^2 \theta)\right\rbrace+d_\chi^2\left\lbrace (s-4m_\chi^2) (1-\cos^2 \theta)\right\rbrace]
\eea
\noindent 

The differential cross section for the process is 
\bea \label{eqn:dsigmadOmega}
\frac{d\sigma}{d\Omega}({ \chi \bar{\chi} \stackrel{\gamma} \longrightarrow e^-e^+} )=\frac{1}{64\pi^2 s}\cdot \left(\sqrt{1-\frac{4m_\chi^2}{s}}\right)^{-1}\cdot\overline{|\mathcal{M}|^2}
\eea
\noindent 

%\bea \label{eqn:dsigmadcostheta}
%\frac{d\sigma}{d(\cos \theta)}=\frac{1}{32\pi s}\cdot\sqrt{1-\frac{4m_\chi^2}{s}}\cdot\overline{|\mathcal{M}|^2}
%\eea
%\noindent 

Finally, the total cross section is given by 
\bea \label{eqn:totalcross-section}
\sigma({\chi \bar{\chi} \stackrel{\gamma} \longrightarrow e^-e^+} )=\frac{\alpha}{6s} \cdot \left(\sqrt{1-\frac{4m_\chi^2}{s}}\right)^{-1} \cdot \left[\mu_\chi^2(s+8m_\chi^2) + d_\chi^2 (s-4m_\chi^2) \right]
\eea
\noindent 
Here $m_\chi$ is the dark matter mass, $\alpha = \frac{e^2}{4 \pi}$ and  
$s=(p_1 + p_2)^2 = (p_3 + p_4)^2$ is the Mandelstam variable.

%%%%%%%%%%%%%%%%%%%%%%%%%%%
\begin{flushleft}
{\bf Thermal averaged cross-section times velocity} 
\end{flushleft}
The thermal averaged cross-section times velocity for the dark matter fermion pair annihilation is given by \cite{kolb, gondolo}
\bea \label{eqn:thermal_averaged_crosssection_times_velocity}
<\sigma_{ \chi \overline{\chi} \rightarrow e^- e^+}~ V_{rel}> = \frac{\int_{m_\chi}^{\infty} \int_{m_\chi}^{\infty} d E_1~d E_2~\sqrt{E_1^2-m_{\chi}^2}~\sqrt{E_2^2-m_{\chi}^2}~(E_1 + E_2)^2~\s_{\chi \overline{\chi} \rightarrow e^- e^+}~f_1~f_2}{\int_{m_\chi}^{\infty} \int_{m_\chi}^{\infty} d E_1~d E_2~E_1~E_2~\sqrt{E_1^2-m_{\chi}^2}~\sqrt{E_2^2-m_{\chi}^2}~f_1~f_2} \nonumber \\
\eea
where the c.m. energy $ E_{c.m.}(=E_1+E_2) =2 E$ (where $E_1 = E_2 = E$) and the relative velocity 
$ V_{rel}=\frac{s}{4E_1 E_2}$. The cross section $\sigma_{\chi \overline{\chi} \rightarrow e^- e^+}$ is given in Eq.~(\ref{eqn:totalcross-section}) and $f_i=1/D_i$ where $D_i = \left(1 + \frac{b}{\tau} E_i  \right)^{\tau} + 1$ with $\rm{i} = 1,2$. Here 
$b=\frac{\beta_0}{4 - 3q}$, $\beta_0 = \frac{1}{k_B T}$ (we are working in the unit where $k_B=1$) and 
$\tau = {1 \over q-1}$. The DM fermion particle and antiparticle number densities 
$n_{\chi}= \int \frac{2 d^3p_1}{(2\pi)^3}~{D_1}^{-1}$ and $n_{\bar{\chi}}= \int \frac{2 d^3p_2}{(2\pi)^3}~{D_2}^{-1}$.
Introducing the dimensionless variables $x_i = E_i/T$~($i=1,2$), we can finally write the Eq.~(\ref{eqn:thermal_averaged_crosssection_times_velocity}) as 
\bea \label{eqn:thermal_averaged_crosssection_times_velocity_final}
<\sigma_{ \chi \overline{\chi} \rightarrow e^- e^+}~ V_{rel}>=\frac{\alpha}{6}\frac{\int_{\frac{m_\chi}{T}}^{\infty} \int_{\frac{m_\chi}{T}}^{\infty} d x_1~d x_2~\sqrt{x_1^2-\frac{m_{\chi}^2}{T^2}}~\sqrt{x_2^2-\frac{m_{\chi}^2}{T^2}}~{\mathcal F}~f_1~f_2}{\int_{\frac{m_\chi}{T}}^{\infty} \int_{\frac{m_\chi}{T}}^{\infty} d x_1~d x_2~x_1~x_2~\sqrt{x_1^2-\frac{m_{\chi}^2}{T^2}}~\sqrt{x_2^2-\frac{m_{\chi}^2}{T^2}}~f_1~f_2}  
\eea
\noindent 
where the function $\mathcal{F}$ is given by
\bea \label{eqn:X}
{\mathcal F}=\left(\sqrt{1-\frac{4m_\chi^2}{T^2(x_1+x_2)^2}} \right)^{-1} \cdot \left[\mu_\chi^2\left\lbrace(x_1+x_2)^2+\frac{8m_\chi^2}{T^2}\right\rbrace + d_\chi^2 \left\lbrace(x_1+x_2)^2-\frac{4m_\chi^2}{T^2}\right\rbrace \right] \nonumber \\
\eea
\noindent 
%\int_{\om_0}^{\infty} d \om \frac{\om (\om^2 - \om_0^2)^{1/2}}{\left(1 + (q-1)b \om \right)^{\tau} -1} $ is the number density of thermal photons or plasmons in the present case ($q$-deformed statistics) which in the normal(undeformed i.e. $q=1$) case takes the form $N_{\g_P}= \frac{1}{\pi^2}
%\int_{\om_0}^{\infty} d \om \frac{\om (\om^2 - \om_0^2)^{1/2}}{e^{\om/T}-1} $. 

Noting the fact that in the $q \to 1$ limit, the $q$-deformed distribution formula gets converted to 
either the  Bose-Einstein or Fermi-Dirac statistical distribution formula (which describes the undeformed scenario)
(see the Appendix for a proof), i.e. 
\bea
f_i(\beta,E_i) = \frac{1}{\left(1 + (q-1)b E_i \right)^{\frac{1}{q-1}} \pm 1} ~\stackrel{q \to 1}{\longrightarrow} \frac{1}{e^{b E_i} \pm 1} \left(= \frac{1}{e^{\beta_0 E_i} \pm 1}\right)
\eea
where $e^{b E_i} = e^{\beta_0 E_i} $ with $b = \frac{\beta_0}{4 - 3 q} = \beta_0$ for $q \to 1$ and 
$\beta_0$ is the inverse equilibrium temperature. Now using Eq.(\ref{eqn:Y0}) we can evaluate $Y_0$ and consequently the relic density for the pair of DM fermions using Eq.(\ref{eqn:relic_density}).
%%%%%%%%%%%%%%%%%%%%%%%%%%%%%%%%%%%%%%%%%%%%%%%%%%%%%%%%%%%%%%%%%%%%%%%%%%%%%%%%%%%%
%%% NUMERICAL ANALYSIS
%%%%%%%%%%%%%%%%%%%%%%%%%%%%%%%%%%%%%%%%%%%%%%%%%%%%%%%%%%%%%%%%%%%%%%%%%%%%%%%%%%%%

%\newpage
\section{Numerical Analysis}
The dark matter produced inside the supernova core via the channel 
$e^- e^+ \to \chi \overline{\chi} $ [Fig. \ref{feynman1}] can contribute to the 
supernova energy loss rate and also similar type of DM fermions(does not depend on the source of production, maybe produced much earlier from other sources as well) contribute to the relic density via the process 
$ \chi \overline{\chi} \to e^- e^+ $ [Fig. \ref{feynman2}].
%, if the emissivity of this channel 
%$\dot{\varepsilon}(e^- e^+ \to \chi \overline{\chi}) \le 10^{19}~\rm{erg~g^{-1}s^{-1}}~(=7.288\times 10^{-27} \rm{GeV})$ 
%and they can contribute to the relic density.
Since the core temperature(T) of the supernova is fluctuating, we follow the $\chi^2$ distribution 
analysis technique \cite{Beck_cohen} here, where the temperature distribution is characterized by 
its mean value  $T (=T_{SN}) = 30~\rm{MeV}$ (see Sec. II B for more details about $\chi^2$ distribution).
%\footnote{Here average temperature of supernova core has been considered as 30 MeV.} 
As mentioned earlier, due to the temperature fluctuation, the ensemble of nucleons, 
electrons, dark matter fermions, photons inside the supernova will tend to follow $q$-deformed or 
Tsallis statistics \cite{Tsallis} (when the deformation parameter $q \neq 1$) 
which is different from the usual Fermi-Dirac and Bose-Einstein statistics 
(when $q=1$). We investigate here the sensitivity of $q$ on the 
dark matter effective scale $\Lda$ for a dark matter fermion of mass($m_\chi$) varying between 
$1 \mbox{-} 100~\rm{MeV}$.

%\newpage 
\subsection{Bound on the effective scale $\Lambda$ from the relic density of DM fermions obtained from 
$\chi \overline{\chi} \stackrel{\gamma}{\longrightarrow} e^{+} + e^{-} $ process}
\begin{figure}[ht!]
\centering
\includegraphics[width=8cm]{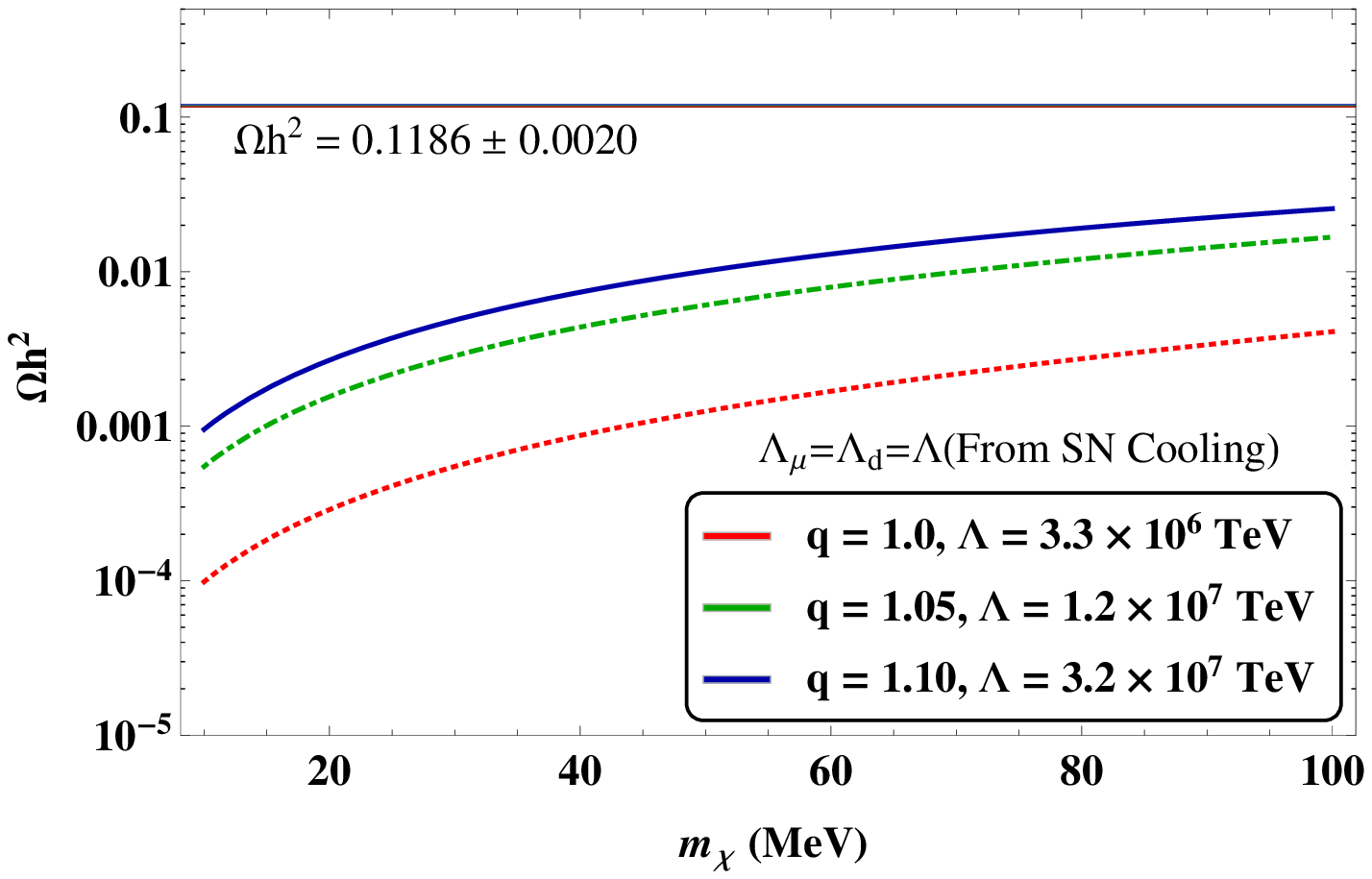}
\caption{ $\Omega h^2$ are plotted against $m_\chi$ 
(in MeV) for $q=1, 1.05~\rm{and}1.1$ for the value of $\Lda$($\Lda_\mu = \Lda_d = \Lda$, Case III) obtained using Raffelt's criteria \cite{Atanu}. }
\protect\label{omega-mdark}
\end{figure}

%We are now to investigate the pair production of light ($\simeq 1 - 30~\rm{MeV}$) dark matter inside the supernova
%core and it's role in the supernova SN1987A cooling process.
Depending on whether the effective coupling of dark matter fermion with photon 
is characterized by a dipole moment operator of magnetic or  electric type, 
there can be three cases as mentioned below \cite{Atanu}:  
\begin{enumerate}
 \item Case I: $\mu_\chi (\sim 1/\Lambda_\mu) \neq 0,~d_\chi (\sim 1/\Lambda_d) = 0$.
 \item Case II: $\mu_\chi (\sim 1/\Lambda_\mu) = 0,~d_\chi (\sim 1/\Lambda_d) \neq 0$.
 \item Case III: $\mu_\chi (\sim 1/\Lambda_\mu) \neq 0,~d_\chi (\sim 1/\Lambda_d) \neq 0$. 
Here $\Lda_\mu = \Lda_d = \Lda$. 
\end{enumerate}
Here $\Lda$ is the scale of the dark matter effective theory. In each of the above three cases, 
we have two possible scenarios corresponding to $q \neq 1$ 
(deformed scenario) and $q = 1$ (undeformed scenario).  \\   
%%%%%%%%%%%%%%%%%%%%%%%%%%%%%%%%%%%%%%%%%%%%%%%%%%%%%%%%%%%%%%%%
\noindent We next calculate the dark matter contribution to the relic density $\Omega h^2$. We obtained consistent results as the contribution is much less than $0.1186$.
In Fig. \ref{omega-mdark}, we have plotted $\Omega h^2$ as a function of the dark matter 
mass $m_\chi$ for different $q$ values corresponding to $\Lda$. The horizontal line corresponds to the 
current experimental bound on the non-baryonic dark matter relic density $\Omega h^2 = 0.1186\pm 0.0020$ (PDG 2017) and the lower set of curves corresponding to different $\Lda$ values (lower bound obtained from the SN1987A cooling  \cite{Atanu}) and $q =1,~1.05$ and $1.1$, respectively. 
%%%%%%%%%%%%%%%%%%%%%%%%%%%%%%%%%%%%%%%%%%%%%%%%%%%%%%%%%%%%%%%%%%%%%%%%%%%%%%%
%%%%%%%%%%%%%%%%%%%%%%%%%%%%%%%%%%%%%%%%%%%%%%%%%%%%%%%%%%%%%%%%%%%%%%%%%%%%%%
%%%%START FROM HERE
\noindent We see that relic density for contribution of a DM fermion of mass $30$ MeV from $5 \times 10^{-4}$ to $5 \times 10^{-3}$ as the deformation 
 parameter $q$ varies from $1.0$  to $1.1$. For a given deformation i.e. $q=1.1$, $\Omega h^2$ varies from 
 $0.001$ to $0.02$ as $m_{\chi}$ increases from $10$ to $100~\rm{MeV}$. This is as per the expectation 
because massive DM fermions should contribute more to the relic density in principle.
%%%%%%%%%%%%%%%%%%%%%%%%%%%%%%%%%%%%%%%%%%%%%%%%%%%%%%%%%%%%%%%%%%%%%%%%%%%%%%%%%%% 
%\noindent Using the non-baryonic matter contribution to relic density  
%$\Omega h^2$ ($ = 0.1186 \pm 0.0020$ from PDG 2017), we find the effective scale 
%$\Lambda$ (combining Eqs. (\ref{eqn:relic_density}) and (\ref{eqn:thermal_averaged_crosssection_times_velociy_final})), which is comparable with the bound obtained by Atanu \etal in} from SN1987A 
%cooling. 
%Raffelt's criterion $\dot{\varepsilon}(e^- e^+ \to \chi \overline{\chi}) \le 10^{19}~\rm{erg~g^{-1}s^{-1}}~(=7.288\times 10^{-27} \rm{GeV})$ and optical depth criterion $\int_{r_0}^{R_c}\frac{dr}{\lambda_{\chi}} \leq \frac{2}{3}$ \cite{Atanu, Dreiner} with $\lambda_{\chi}=\frac{1}{n_e \cdot \sigma_{e \chi \rightarrow e \chi}}$ \cite{Atanu} and $r_0=0.9 R_c$, where $R_c  \simeq 10 $ km.
%%%%%%%%%%%% 

\begin{figure}[htb]
  \centering
   \subfloat[$ q=1.0 $]{\includegraphics[width=0.49\textwidth]{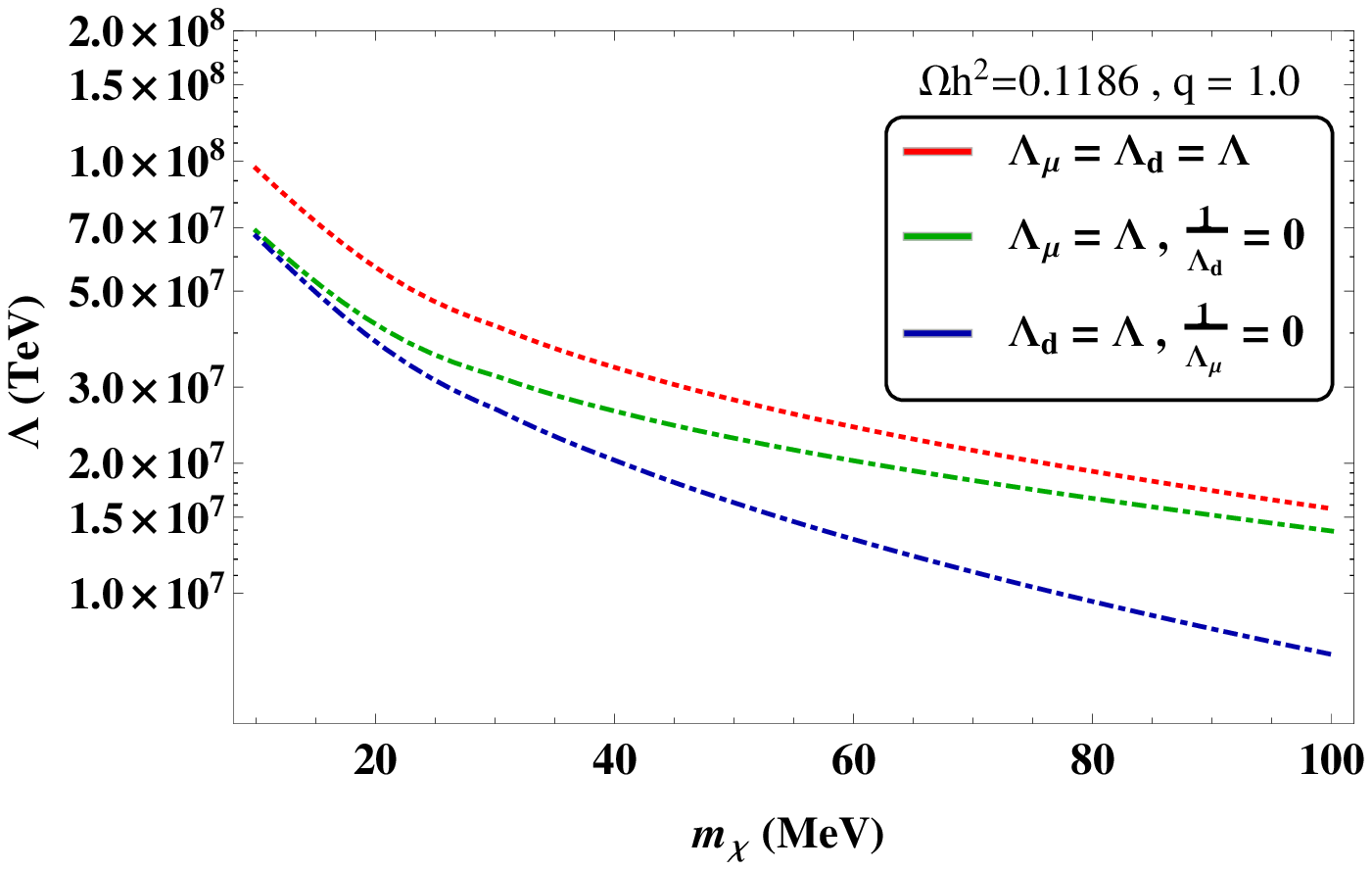}\label{Lda-mdarkq1}}
  \hfill
  \subfloat[$ q=1.1 $]{\includegraphics[width=0.49\textwidth]{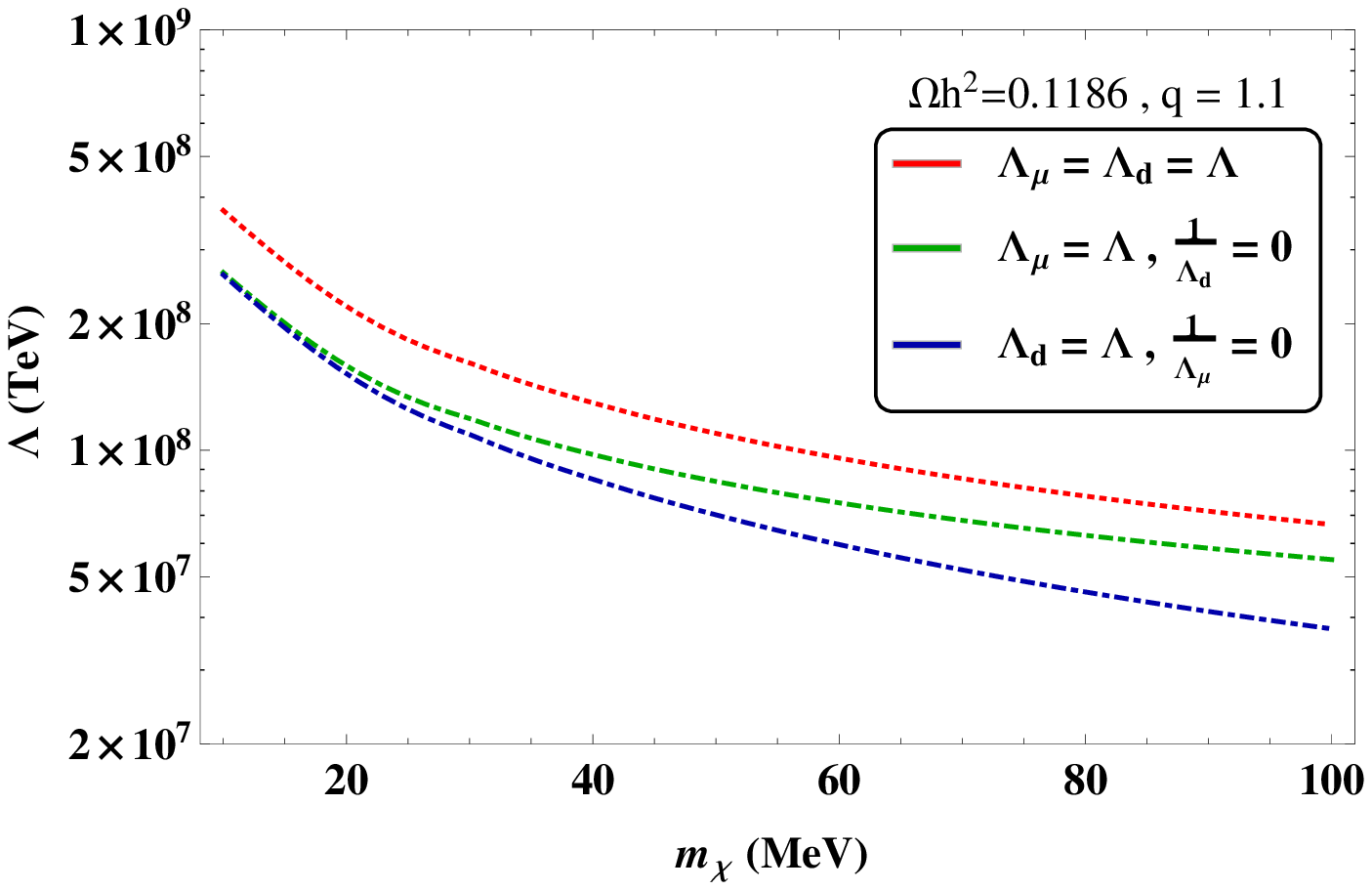}\label{Lda-mdarkq2}}
\caption{$\Lda$ (in TeV) is plotted against $m_\chi$ (in MeV) for  $q=1~\rm{and}~q=1.1$.}
\protect\label{Lda-mdarkq}
\end{figure}

\begin{figure}[htb]
\centering
\includegraphics[width=8.0cm]{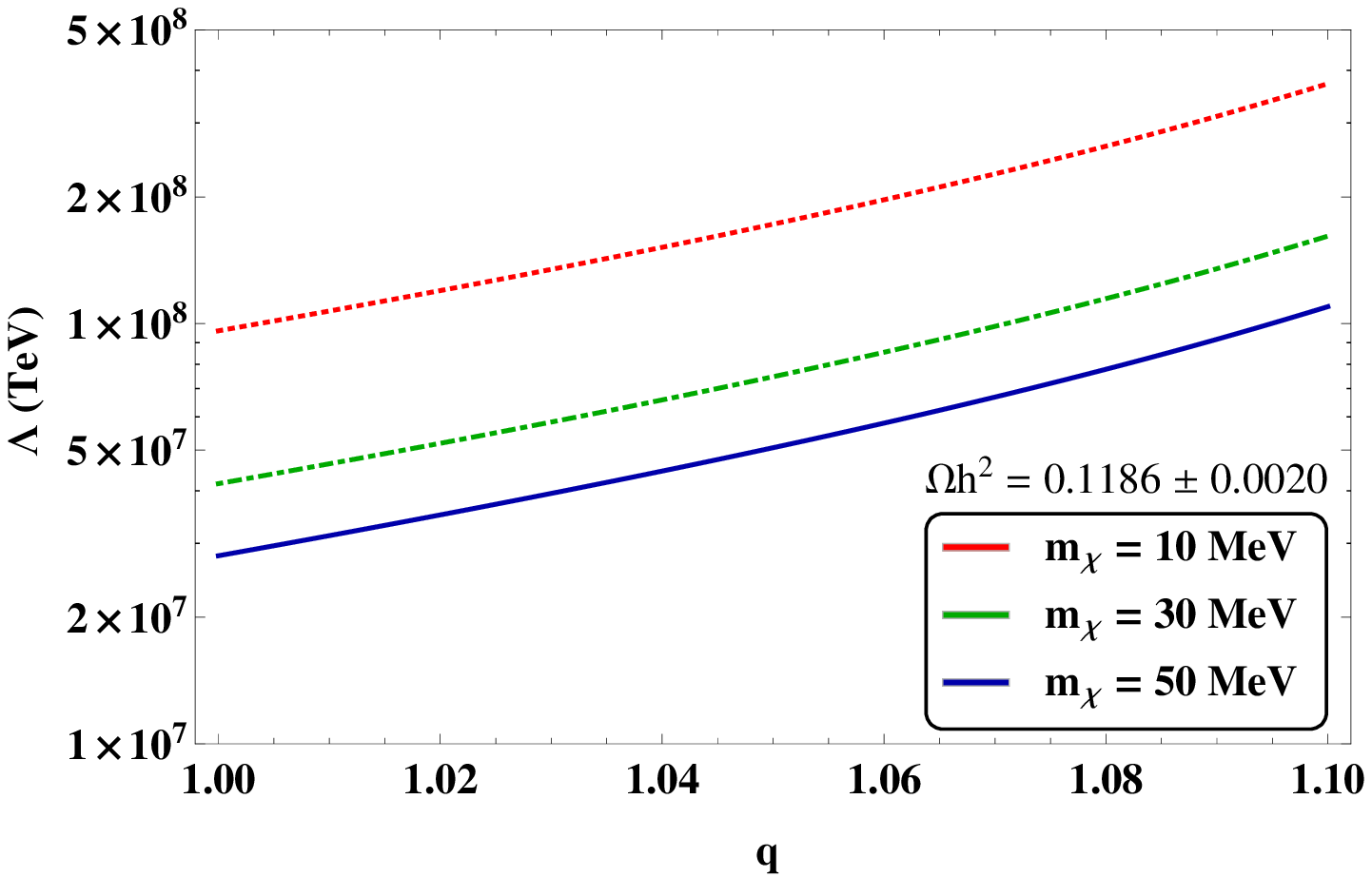}
\caption{$\Lda$ (in TeV) is plotted against $q$ for different $m_\chi$ and $\Omega h^2=0.1186 \pm 0.0020$. }
\protect\label{lda-qparam}
\end{figure}

\begin{figure}[htb]
\centering
 \subfloat[$ q=1.0 $]{\includegraphics[width=0.49\textwidth]{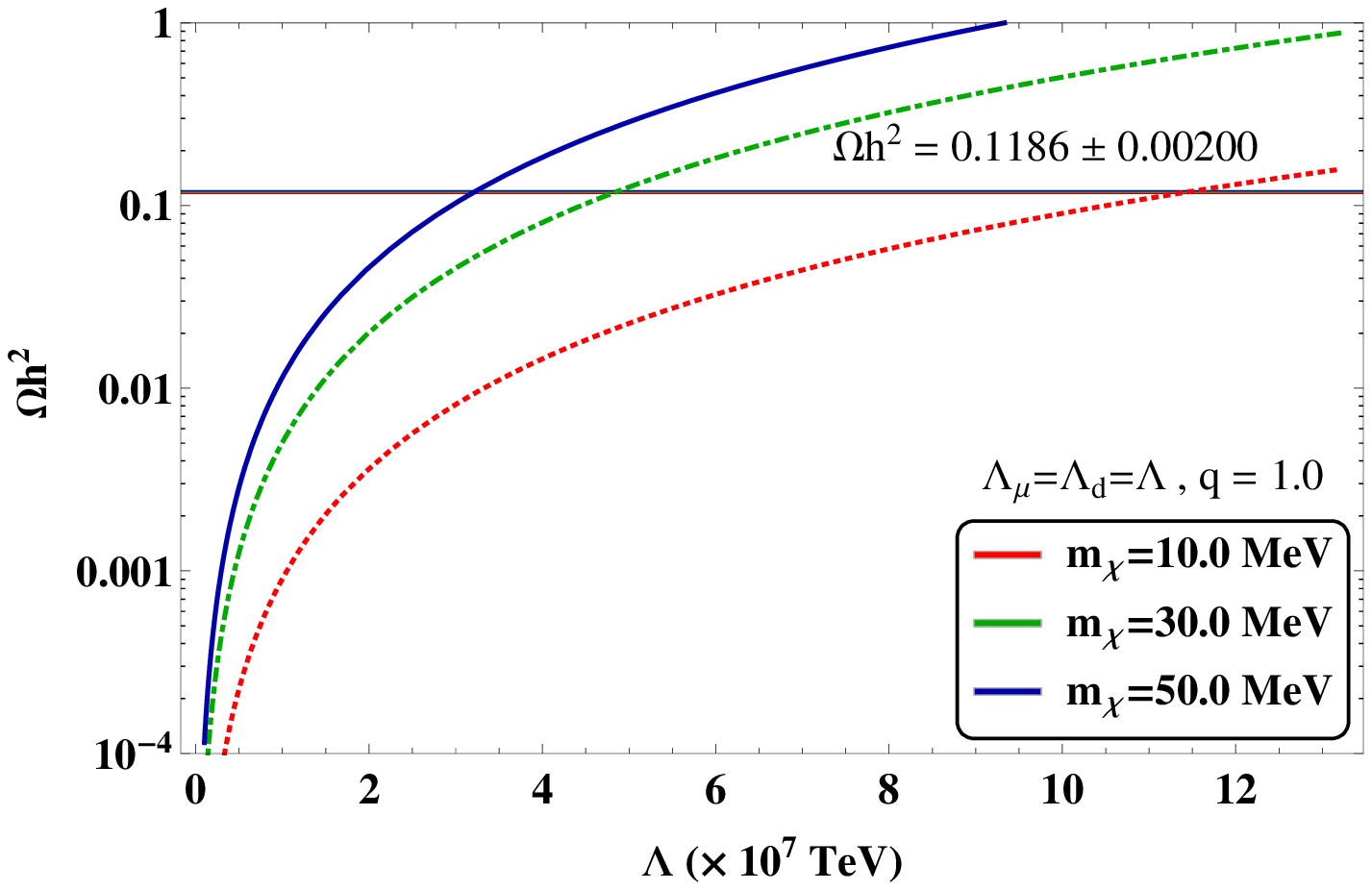}\label{omega-lambda2}}
 \hfill
  \subfloat[$ q=1.1 $]{\includegraphics[width=0.49\textwidth]{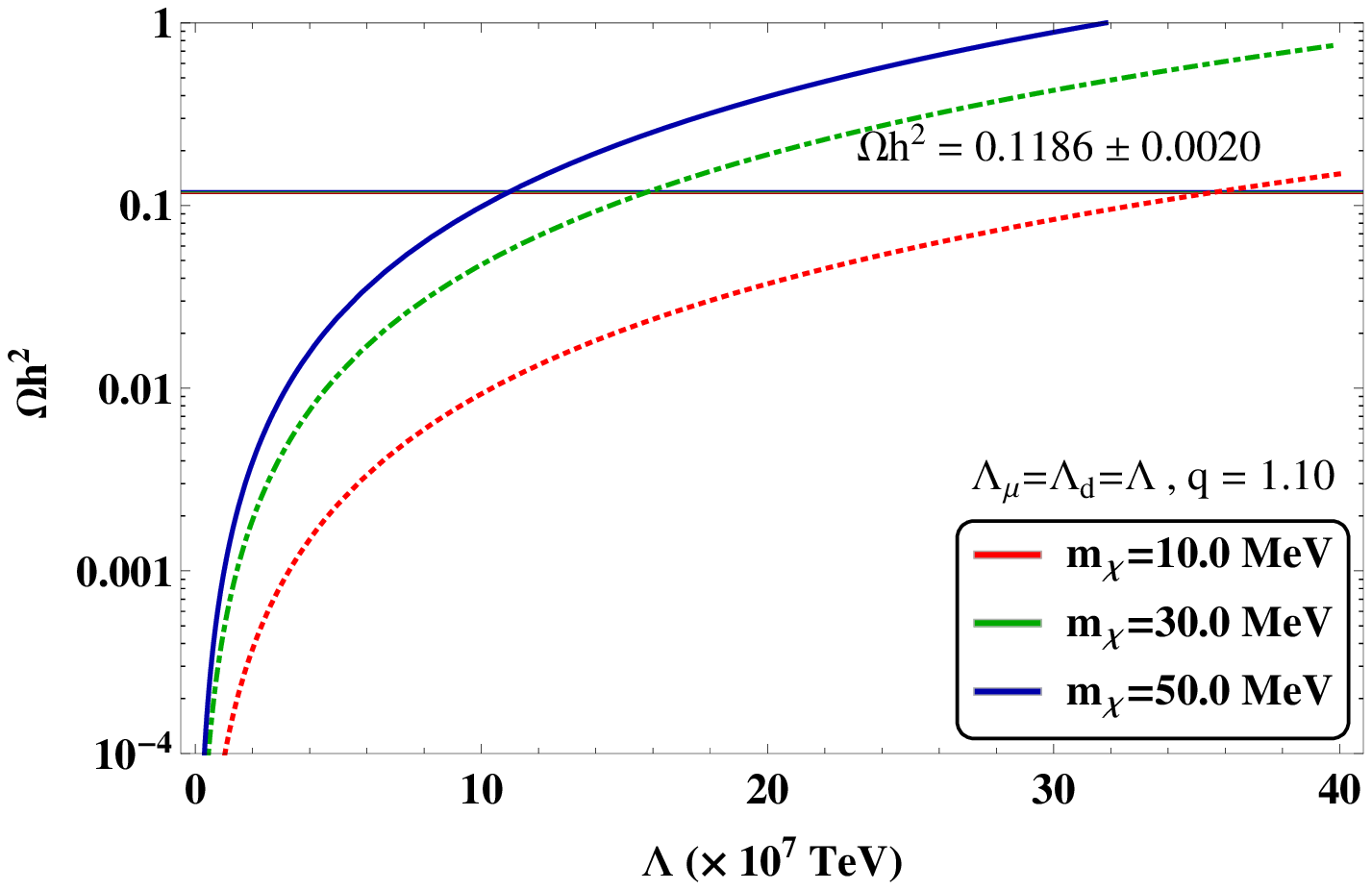}\label{omega-lambda1}}
\caption{\it $\Omega h^2$ is plotted against $\Lda$ (in TeV) for $q=1$ and $q=1.1$ and different $m_\chi$ values.}
\label{omega-lambda}
\end{figure}

\noindent In Fig. \ref{Lda-mdarkq}, we have plotted the upper bound $\Lda$ as a function of 
$m_\chi$ for $q=1$(Fig. \ref{Lda-mdarkq1}) and $q=1.1$(Fig. \ref{Lda-mdarkq2}) corresponding to 
$\Omega h^2 = 0.1186$ for three cases. 
%%%%%%%%%%%%%%%%%%%%%%%%%%%%%%
%%%%%%%%%%%%%%%%%%%%%%%%%%%%%%
\noindent 
For a given $q$, the upper bound on $\Lda$ decreases as $m_\chi$ increases. As an example, for $q=1.0$, we 
see that as $m_\chi$ increases from $10$ to $100~\rm{MeV}$, $\Lda$(Case III) decreases from $1 \times 10^8~\rm{TeV}$ to $1.6 \times 10^7~\rm{TeV}$. On the right, the same is shown for $q=1.1$, where $\Lda$ changes from 
$\sim 4 \times 10^8~\rm{TeV}$ to $\sim 7 \times 10^7~\rm{TeV}$ for the same $m_\chi$ range. 
In Fig. \ref{lda-qparam}  we have plotted the upper bound on $\Lda$ (obtained from the 
relic density constraint) as a function of $q$ corresponding 
to  $m_\chi =10(\rm{topmost}),~30~(\rm{middle})$ and $50~(\rm{lowermost})$ MeV (for Case III) using 
$\Omega h^2=0.1186,~0.1206$ and $0.1166$, respectively. 
%%%%%%%%%%%%%%%%%%
%%%%%%%%%%%%%%%%%%%%%%%
% We find that $\Lda$ changes from $1 \times 10^8~\rm{TeV}$ to $4 \times 10^8~\rm{TeV}$
% as $q$ changes from $1.0$  to $1.1$ for a $10~\rm{MeV}$ mass DM fermion. 
For a given $m_\chi$ all the three curves corresponding to three different values of $\Omega h^2$ are found to be almost overlapping.
The following observations are in order: (a) For a given $m_\chi$, the upper bound on $\Lda$ increases with the increase in $q$. 
 As an example, for $m_\chi = 30~\rm{MeV}$, $\Lda$ changes from $4 \times 10^7~\rm{TeV}$ to $1.6 \times 10^8~\rm{TeV}$ as $q$ changes from $1.0$ 
 to $1.1$ and (b) $\Lda$ decreases with the increase in $m_\chi$ for a particular $q$. As an example, 
 for $q=1.1$, we see that  $\Lda$ decreases from $4 \times 10^8~\rm{TeV}$ to $1.1 \times 10^8~\rm{TeV}$ 
 as $m_\chi$ increases from $10$ to $50~\rm{MeV}$.
%\hspace*{-0.5in}
%%%%%%%%%%%%%%%%%
In Fig. \ref{omega-lambda}, we have ploted $\Omega h^2$ against the upper bound on $\Lda$ (in TeV) for 
different mass fermion dark matter fermion for Case III in $q$-deformed 
(Fig. \ref{omega-lambda2}) and undeformed (Fig. \ref{omega-lambda1}) scenarios. In both figures the region below the 
horizontal lines $\Omega h^2=0.1186\pm 0.0020$ are allowed as this is the experimental 
upper bound on value of total non-baryonic contribution to the relic density.
%%%%%%%%%%%%%%%%%
%%%%%%%%%%%%%%%%%%%%%%%%%%%%%%%%%%%%%%% 
So theoretically estimated relic density for a particular particle produced in a particular process should 
not exceed the value $\Omega h^2=0.1186\pm 0.0020$, but can be less than or in extreme case 
equal to this. Fig. \ref{omega-lambda} suggests that for a given mass of DM fermions 
$m_{\chi}$ an increment in the value of the effective scale $\Lda$ will cause a significant
increment in the contribution to the relic density for that DM fermions. This observation is
physically consistent as well. As $\Lda$ increases, the coupling constants($\mu, d$) becomes 
weaker. As a result the DM couples more weakly to the concerned standard 
model(SM) paricles(here $e^+~\rm{and}~e^-$) and hence after their production, they are more likely to contribute to the relic density, rather than interacting with SM particles and gets scattered.
%%%%%%%%%%%%%%%%%%%%%%%%%
\subsection{Discussion of the lower bound obtained from SN1987A Cooling(Raffelt's  criteria), 
Free Streaming(Optical depth criteria) with the upper bound obtained from the Relic density contribution} 
\begin{figure}[htb]
   \centering
  \subfloat[$ q=1.0 $]{\includegraphics[width=0.49\textwidth]{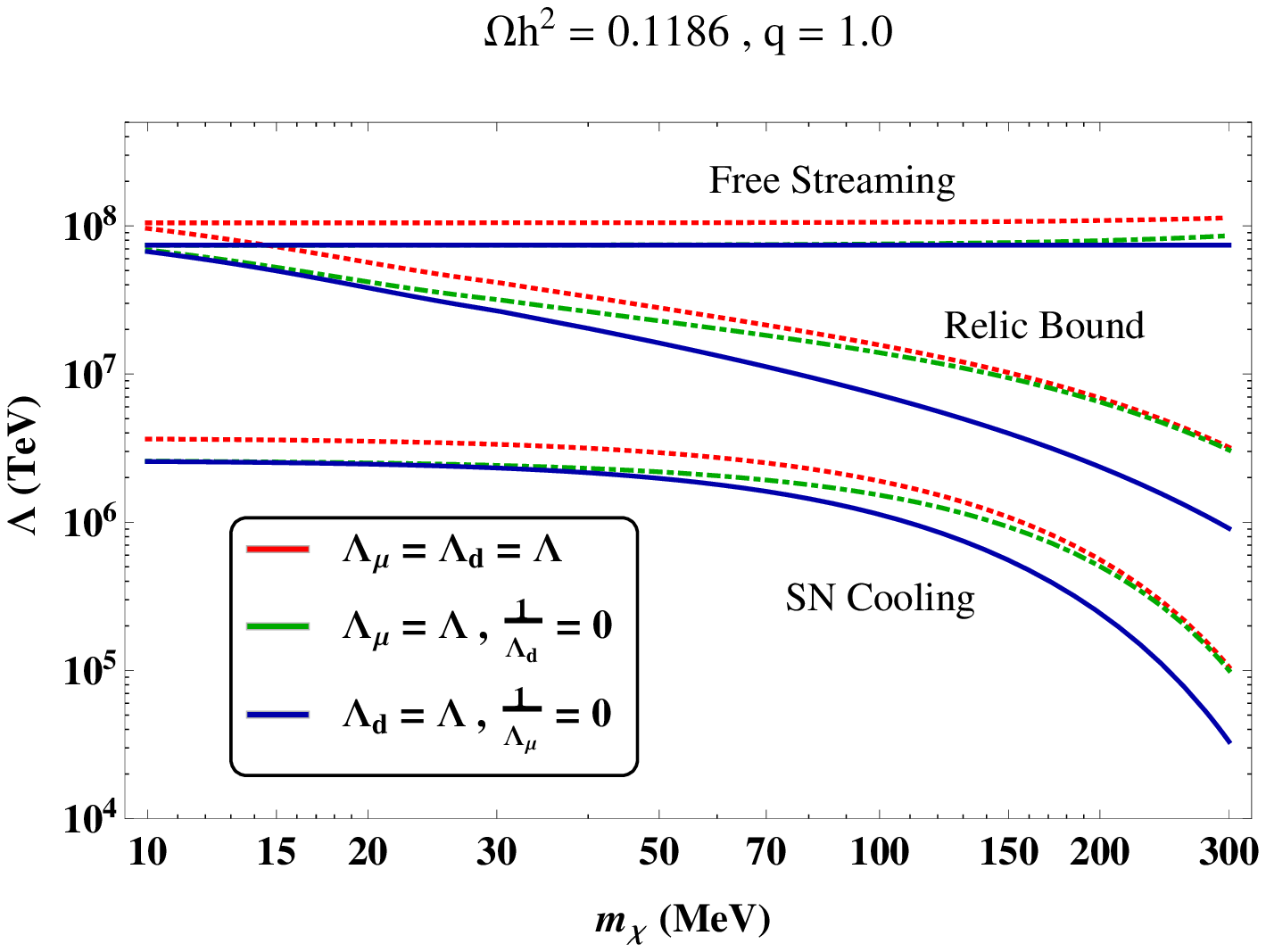}\label{lambda-mchi-sn-fs-relic1}}
   \hfill
   \subfloat[$ q=1.1 $]{\includegraphics[width=0.49\textwidth]{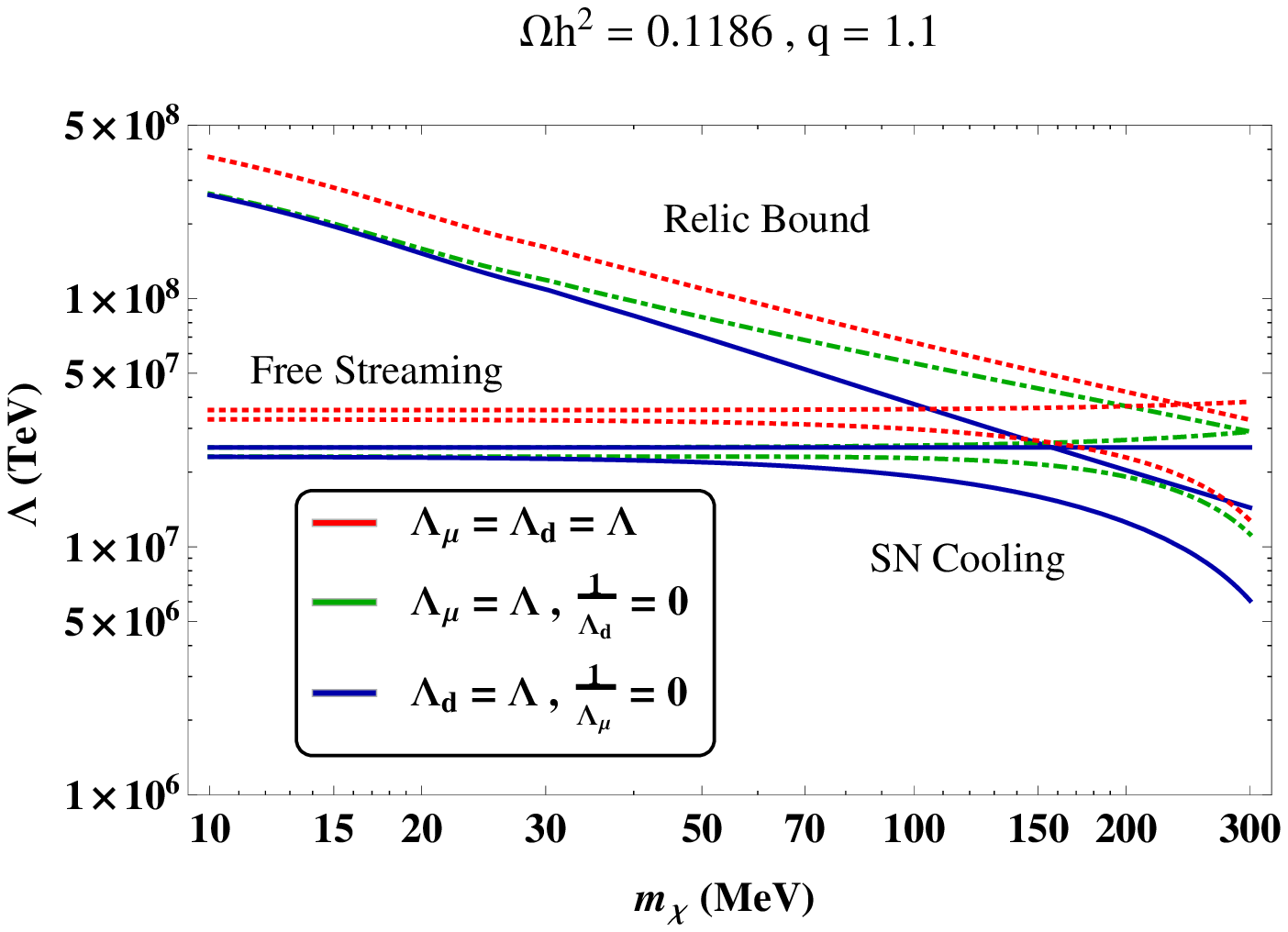}\label{lambda-mchi-sn-fs-relic2}}
\caption{\it $\Lda$(in TeV) (obtained from SN1987A cooling, free streaming \cite{Atanu} and 
using Relic bound)  is plotted against $m_{\chi}$(in MeV) for undeformed ($q=1$)
and $q$-deformed scenario.}
\label{lambda-mchi-sn-fs-relic}
\end{figure}

\noindent We find that the lower bound on $\Lda$ obtained from the Raffelt's criteria(SN1987A 
cooling), Optical depth criteria(free streaming of DM fermion) and the upper bound on $\Lda$ obtained from the relic density contribution, 
obtained in both deformed and un-deformed scenarios are consistent with each other. The lower bound obtained on $\Lambda_\mu$ in undeformed 
scenario ($q=1.0$) is comparable with that obtained by Kadota \etal \cite{KS}. Below in 
Table I, we make a comparative study  of the lower and upper bounds on $\Lambda$ obtained by using the above three 
criteria. \\
%%%%%%%%%%%%%%%%%%%%
 \noindent {{\textbf{Table I}}: {The bound on the effective scale 
 $\Lda=\Lda_{\mu}=\Lda_{d}$ (TeV) (lower bound obtained from optical depth criterion, Raffelt's criteria
and upper bound obtained from relic density experimental value) are shown for different 
 dark matter mass $m_\chi$ (MeV) in the undeformed($q=1$) and $q$-deformed scenario, for free streaming(optical depth criterion) there is a range of the reported value \cite{Atanu} as it depends on the other supernova properties like temperature, densities etc. \cite{Dreiner, DHLP, Debades, Das, Ellis, Lau, Sathees, Sumiyoshi, Hagel, FHS}}}
\begin{center}
\begin{tabular}{m{2cm} m{3cm} m{4cm} m{3cm} m{3cm} }
\hline \hline
 & $m_\chi~(\rm{MeV})$ & \multicolumn{3}{c}{$\Lda=\Lda_\mu=\Lda_d~\rm{(TeV)}$}  \\
\cline{3-5}
  &   &  Free streaming & SN cooling & Relic bound \\
\hline
 $q=1.0$  & $10$ & $\sim (3.6$-$10.5) \times 10^7$ & $3.66\times 10^6$ & $1.16 \times 10^8$ \\

   & $30$ & $\sim (3.6$-$10.5) \times 10^7$ & $3.34\times 10^6$ & $4.92 \times 10^7$ \\

   & $50$ & $\sim (3.6$-$10.5) \times 10^7$ & $3.00\times 10^6$ & $3.22 \times 10^7$ \\ 
\hline
 $q=1.05$  & $10$ & $\sim (3.6$-$10.5) \times 10^7$ & $1.25\times 10^7$ & $1.76 \times 10^8$ \\

   & $30$ & $\sim (3.6$-$10.5) \times 10^7$ & $1.2\times 10^7$ & $5.37 \times 10^7$ \\

   & $50$ & $\sim (3.6$-$10.5) \times 10^7$ & $1.15\times 10^7$ & $7.73 \times 10^7$ \\ 
\hline
 $q=1.1$  & $10$ & $\sim (3.6$-$10.5) \times 10^7$ & $3.26\times 10^7$ & $3.53 \times 10^8$ \\

   & $30$ & $\sim (3.6$-$10.5) \times 10^7$ & $3.23\times 10^7$ & $1.58 \times 10^8$ \\

   & $50$ & $\sim (3.6$-$10.5) \times 10^7$ & $3.2\times 10^6$ & $1.11 \times 10^8$ \\ 
         
\hline \hline
\end{tabular}
\end{center}
%%%%%%%%%%%%%%%%%%%%%%%%%%%%%
% \noindent {{\textbf{Table I}}: {\it The lower bound on the effective scale $\Lda=\Lda_{\mu}=\Lda_{d}$ (TeV) (obtained from Raffelt's criteria, optical depth criterion and relic density experimental value) are shown for different 
% dark matter mass $m_\chi$ (MeV) in the undeformed and $q$-deformed scenario.}} \\
%%%%%%%%%%%%%%%%%%%
In Fig. \ref{lambda-mchi-sn-fs-relic}, we have shown the bounds on $\Lda$ (in TeV) against the 
DM mass $m_{\chi}$ (in MeV) obtained from SN1987A energy loss rate
(Raffelt's criteria), Optical depth criterion(free streaming) and using Relic density of 
non-baryonic matter $\Omega h^2 = 0.1186$ (PDG 2017). We obtain lower bound from Raffelt's criteria and Optical depth criterion and on the other hand we obtain upper bound using the maximum possible relic density of non-baryonic matter. Clearly, the region between the SN cooling and relic bound curves are allowed, forbidden regions are located below the SN cooling curve and above the relic curve. On the left the plots are shown for 
undeformed $q=1$ scenario, while on the right the plots are shown for $q=1.1$.
%%%%%%%%%%%%%%%%%%%%%%%%%
%%%%%%%%%%%%%%%%%%%%%%%%%%%%%%%%%%%%%%%%%%%%%%%%%%%%%%%%%%%%%%%%%%%%%%%%%%%%%%%%%%
In Fig. \ref{lambda-mchi-sn-fs-relic1}($q=1$ undeformed), we see that there is a widespread region between the lower bound on $\Lda$ obtained from the SN1987A cooling 
criteria and the upper bound on $\Lda$ obtained from the maximum possible value of relic density for an ultra-light mass $m_\chi = 10$ MeV 
DM fermion, however the lower bound obtained from the free streaming criteria lies in the forbidden region. 
In the $q$-deformed scenario (with $q=1.1$) the story is slightly different(Fig. \ref{lambda-mchi-sn-fs-relic2}).  
For a light mass dark matter (with $m_\chi$ ranging between $10$ MeV to $100$ MeV), the lower bounds obtained on 
$\Lda$ from SN1987A cooling and free streaming criteria  more-or-less agree with each 
other, but the upper bound differs by an order of magnitude obtained from the relic density constraint. For a heavier 
DM fermion with mass $m_\chi \sim 100 ~-~ 200$ MeV, the lower and upper bounds obtained from the supernovae 
cooling, free streaming criteria and relic density, are found to be comparable \cite{Atanu2}. This restricts or precisely forbids the production of heavier dark matter fermions inside supernova core as this is forbidden by the upper bound obtained from the relic density constraints.
%i.e., $q$-deformed (Fig. \ref{lambda-mchi-sn-fs-relic2}) and undeformed (Fig. \ref{lambda-mchi-sn-fs-relic1}).
%%%%%%%%%%%%%%
\section{Conclusion}
The dark matter fermion produced inside the supernova SN1987A core in electron-positron collision 
$e^+ e^- \to \chi \overline{\chi}$ , can take away the energy released in the supernova 
explosion and similar kind of dark matter fermion produced from some other sources as well can contribute to the relic density.
%Working within the formalism of $q$-deformed statistics we evaluate the contribution of the produced dark matter fermions to the relic density. 
Working within the formalism of $q$-deformed statistics we find the DM contribution 
to the relic density and using the experimental bound on the relic density (of the cold 
non-baryonic matter) $\Omega h^2 = 0.1186 \pm 0.0020$ (obtained from the measurement of the anisotropy of the cosmic microwave background (CMB) and of the 
spatial distribution of galaxies), we obtain a upper bound on the effective scale $\Lambda$. 
In the un-deformed(deformed) scenario $q=1$($q=1.1$), for a light mass ($m_\chi = 30$ MeV) dark matter, we   
find the upper bound on $\Lambda = 4.9 \times 10^7$ TeV($\Lambda = 1.6 \times 10^8$ TeV) from the relic density. 
This is consistent with the lower bound $3.3 \times 10^6$ TeV($\Lambda = 3.2 \times 10^7$ TeV) obtained from the Raffelt's criteria on the supernovae energy 
loss rate and $3.6 \times 10^7$ TeV($\Lambda = 3.6 \times 10^7$ TeV) obtained from the optical 
depth criteria on the free streaming of the dark matter fermion in the un-deformed(deformed) 
scenario with $q=1$($q=1.1$), respectively. Here we can note one interesting fact that, for $q$-deformed scenario with $q=1.1$ both the lower bound curves of $\Lda$ lies below the upper bound curves due to the relic density consideration which is physically consistent unlike the undeformed one($q=1.0$).

\appendix
\section{From $q$-deformed statistics to undeformed scenario}

 In general, the distribution function for the $q$-deformed statistics is \cite{Tsallis}
\bea 
D_i=\left(1+\frac{b}{\tau}(E_i-\mu_i)\right)^{\tau}+1 
\eea
with $b=\frac{\beta_0}{4-3q}$, $\beta_0=\frac{1}{k_B T}$ (we work in the unit $k_B=1$) and $\tau=\frac{1}{q-1}$.

 In terms of the dimensionless quantity $x_i=\frac{E_i}{T}$
\bea
D_i=\left(1+b (q-1)(T x_i-\mu_i)\right)^{\frac{1}{q-1}}+1
\eea
 
 Now replacing $q-1$ by $m$, ($m \rightarrow 0$ as $q \rightarrow 1 $)
\bea
\left(1+b (q-1)(T x_i-\mu_i)\right)^{\frac{1}{q-1}}=\left(1+b m(T x_i-\mu_i)\right)^{\frac{1}{m}}=y \mathrm{(say)}
\eea
Now
\bea
\lim_{m\to 0} y &=& \lim_{m\to 0} \left(1+b m(T x_i-\mu_i)\right)^{\frac{1}{m}} \nonumber \\
\implies \lim_{m\to 0} \ln y &=& \lim_{m\to 0} {\frac{1}{m}} \ln \left(1+b m(T x_i-\mu_i)\right)
\nonumber \\
&=& \lim_{m\to 0} \frac{1}{1+b m(T x_i-\mu_i)} b (T x_i-\mu_i)
\nonumber \\
&=& b (T x_i-\mu_i) \nonumber
\eea
 Also for $q \rightarrow 1 $, we find $b(=\frac{\beta_0}{4-3q})=\beta_0=\frac{1}{k_B T}=\frac{1}{T}$. So
we find 
 \bea
 \lim_{q\to 1} \ln y &=& \beta_0 (T x_i-\mu_i) 
 \nonumber \\
\implies \lim_{q\to 1} y &=& \exp \left[\frac{1}{T} (T x_i-\mu_i) \right]
 \nonumber \\
 &=& \exp \left[x_i- \frac{\mu_i}{T} \right] \nonumber
 \eea
 Clearly, in the undeformed scenario (i.e. $q=1 $)
 \bea
 \lim_{q\to 1} D_i = \lim_{q\to 1} y +1 = \exp \left[x_i- \frac{\mu_i}{T} \right] +1 ~~[Proved] 
 \eea

\subsection{Feynman rules}
\label{app:feyn}
%%%%%%%%%%%%%%%%%%%%%%%%%%%%
Process: $e^- e^+ \stackrel{\gamma}{\longrightarrow}  \chi \overline{\chi}$: 

{  $e^-~e^+ \to \gamma$ ~vertex: } $ie \gamma^\mu$ 

{  $\gamma \to \chi~\overline{\chi}$ ~vertex: } $i\left( \mu_\chi \sigma^{\mu \nu} q_\nu 
                                               + d_\chi \sigma^{\mu \nu} q_\nu \gamma^5 \right)$ 
%%%%%%%%%%%%%%%%%%%%%%%%%%%%%%
\acknowledgments

\noindent The authors would like to thank Dr. Selvaganapathy J (PRL, Ahmedabad, India) and Dr. Debasish Majumdar (SINP, Kolkata, India)
for useful discussions. Also the authors are very much thankful to Dr. Bhupal Dev (Department of Physics and McDonnell Center for the Space Sciences, Washington University,
St. Louis, MO 63130, USA) for effective suggestions. This work is partially funded by the SERB, Government of India, Grant No. EMR/2016/002651.

%%%%%%%%%%%%%%%%%%%%%%%%%%%%%%

%%%%%%%%%%%%%%%%%%%%%%%%
\end{document}